%% file: STEREO_helaine.tex
\documentclass[12pt]{article}
\usepackage{graphicx}

\usepackage{subfig}

\newcommand\pubnumber{}
\newcommand\pubdate{\today}

\def\gre{LPSC, Universit\'e Grenoble Alpes, CNRS/IN2P3, FRANCE}
\def\support{\footnote{Work supported by the ANR grant 13-BS05-0007.}}

\def\Title#1{\begin{center} {\Large #1 } \end{center}}
\def\Author#1{\begin{center}{ \sc #1} \end{center}}
\def\Address#1{\begin{center}{ \it #1} \end{center}}

\newcommand\pubblock{\rightline{\begin{tabular}{l} \pubnumber\\
         \pubdate  \end{tabular}}}
\newenvironment{Abstract}{\begin{quotation}  }{\end{quotation}}
\newenvironment{Presented}{\begin{quotation} \begin{center} 
             PRESENTED AT\end{center}\bigskip 
      \begin{center}\begin{large}}{\end{large}\end{center} \end{quotation}}

\input econfmacros.tex

\begin{document}
\begin{titlepage}
\pubblock

\vfill
\Title{Sterile neutrino search at the ILL nuclear reactor: the STEREO experiment}
\vfill
\Author{ Victor H\'elaine for the STEREO collaboration\support}
\Address{\gre}
\vfill
\begin{Abstract}
Search for a light sterile neutrino is currently a hot topic of neutrino physics, arising from the so-called gallium and reactor anomalies, in which a deficit of neutrinos was observed with respect to expectations. Such anomalies could be explained by short distance oscillations towards a sterile state, with $\Delta \mathrm{m}^2\sim$1\,eV$^2$. The STEREO detector has been designed to track the electron anti-neutrino energy spectrum distortion from 3 to 8\,MeV due to such a new $L/E$ oscillation, and should therefore confirm or reject the light sterile neutrino hypothesis. Electron anti-neutrinos produced by the compact reactor core of the Institut Laue-Langevin (ILL) will be detected in a 6-cells segmented volume of Gd-loaded liquid scintillator through the inverse $\beta$-decay process. The STEREO detector is being set-up and will be commissioned in fall 2016, and start data taking soon after. In this paper we will present the final design of the detector and its status, as well as its expected sensitivity.
\end{Abstract}
\vfill
\begin{Presented}
 NuPhys2015, Prospects in Neutrino Physics\\
Barbican Centre, London, UK,  December 16--18, 2015
\end{Presented}
\vfill
\end{titlepage}
\def\thefootnote{\fnsymbol{footnote}}
\setcounter{footnote}{0}

\section{Introduction}
\label{sec:Introduction}

Possibly associated to previous anomalies observed in neutrino radioactive source experiments (summarized in \cite{Giunti2011}) and accelerator based neutrino experiments \cite{Aguilar2001}, the 3 to 6\,\% increase of the revised electron anti-neutrino ($\bar{\nu_e}$) flux predictions for different reactor oscillation experiments \cite{Mueller2011,Mention2011} re-triggered the quest for a sterile neutrino. Indeed, the reactor $\bar{\nu_e}$ counting rates predicted in the 3-neutrino picture became inconsistent with experimental data at about 3$\sigma$, showing a detected neutrino deficit of about 7\,\%. 

This reactor anti-neutrino anomaly (RAA) could be explained by a new neutrino oscillation towards a sterile state -- not interacting weakly -- at short baseline. In that case, the survival probability $P_{\bar{\nu_e}\rightarrow\bar{\nu_e}}$ for a  $\bar{\nu_e}$ with an energy $E_{\bar{\nu_e}}$ after a propagation over a distance $L$ is approximated by
\begin{equation}
P_{\bar{\nu_e}\rightarrow\bar{\nu_e}}(E_{\bar{\nu_e}},L)=1-\sin^2{2\theta_{\mathrm{new}}	\sin^2{\left(1.27\frac{\Delta m^2_{\mathrm{new}}L}{E_{\bar{\nu_e}}}[\mathrm{eV}^2 ] [\mathrm{m}] / [\mathrm{MeV}])\right)}}
\label{eq:osc}
\end{equation}
where the best fit oscillation parameters coming from the RAA are $\Delta m^2_{\mathrm{new}}\simeq2.3\,$eV$^2$ and $\sin^2{2\theta_{\mathrm{new}}}\simeq0.13$. The existence of such a sterile neutrino would be a major discovery in particle physics.

However, an error in nuclear reactor $\bar{\nu_e}$ flux predictions or an underestimation of errors are not excluded as argued for example in \cite{Vogel2016}. Therefore, the goal of the STEREO experiment is to probe the parameter space region allowed by the RAA using very short baseline reactor neutrinos as shown in Fig.~\ref{fig:STEREOContour}.

\section{Principle of the experiment and detector design}
\label{sec:Principle of the experiment and detector design}

In order to probe the existence of light sterile neutrinos, the strategy of STEREO is to measure with exquisite precision the evolution of the $\bar{\nu_e}$ energy spectrum at short distance (10\,m) from the compact reactor core of the Institut Laue-Langevin (ILL), without using $\bar{\nu_e}$ flux normalization to avoid any ambiguity.

According to Eq.~\ref{eq:osc}, the pattern arising from active to sterile neutrino oscillations depends both on the $\bar{\nu_e}$ energy and on the distance travelled from the core to the detection vertex. The distance between the oscillation extrema is of about 2\,m for reactor neutrinos. Therefore, the detector is 2.2\,m long and is segmented in 6 identical cells to measure the relative distortions of the  $\bar{\nu_e}$ energy spectrum between them.

\begin{figure}[htb]
\centering
\subfloat[]{\label{fig:STEREOContour}\includegraphics[width=0.4\linewidth]{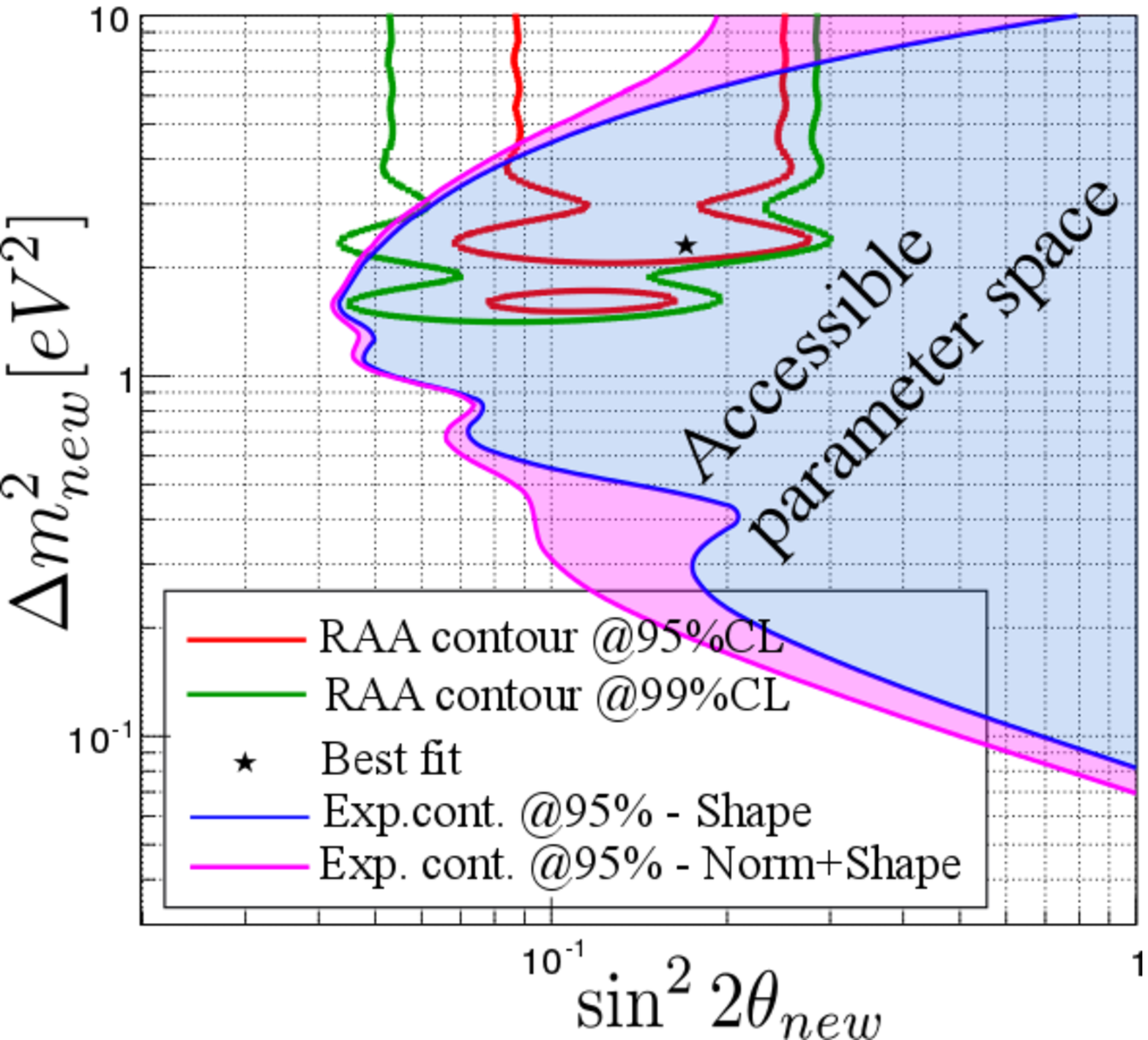}}
\subfloat[]{\label{fig:STEREODetector}\includegraphics[width=0.55\linewidth]{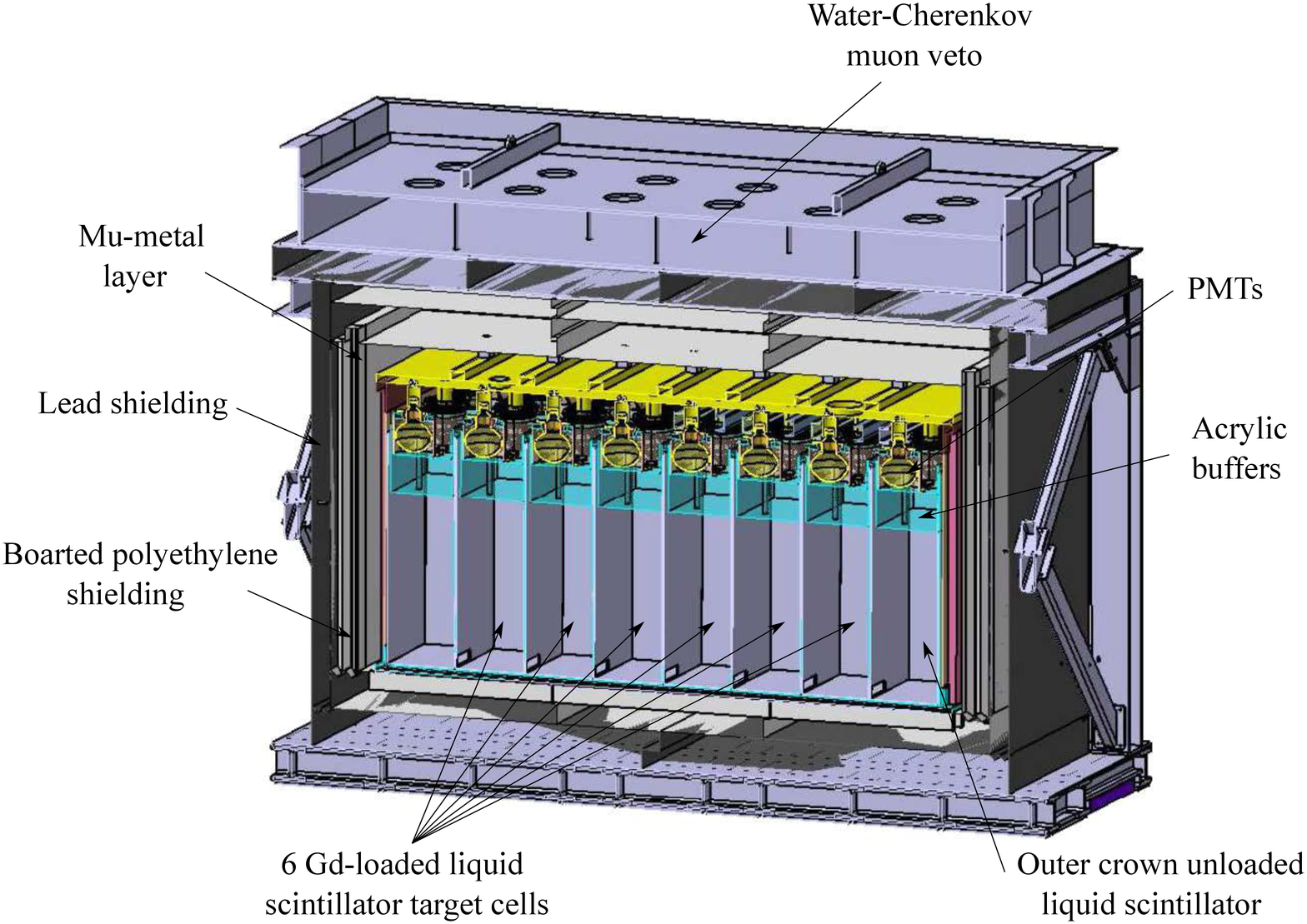}}
\caption{Left panel: the expected sensitivity contours of the STEREO experiment after 2 years of data taking, in the parameter space $\left(\Delta m_{\mathrm{new}}^2, \sin^2\theta_{\mathrm{new}}\right)$, cover the reactor anti-neutrino anomaly. Right panel: cut view of the STEREO detector.}
\end{figure}

The process used to detect $\bar{\nu_e}$ events is the so-called inverse $\beta$-decay (IBD), depicted in Eq.~\ref{eq:ibd}. 
\begin{equation}
\bar{\nu_e}+p\rightarrow e^++n
\label{eq:ibd}
\end{equation}
The signature of such a process in the Gd-loaded liquid scintillator used in the STEREO detector (see Fig\,\ref{fig:STEREODetector}) is a prompt signal from the positron energy deposit, detected in delayed coincidence with the 8\,MeV $\gamma$-cascade produced by the neutron capture on a Gd nucleus The light produced in the target is collected by four 8'' photomultipliers (PMT) per cell via an acrylic buffer. The 6 cells (total volume 2\,m$^3$) are used for a passive vertex determination. The cell thickness of 40$\,$cm is similar to the size of the ILL reactor core, in order to avoid any oscillation smearing.

The target is surrounded by a 40\,cm thick outer crown filled with unloaded liquid scintillator (3$\,$m$^3$) in order to detect escape gammas. This is used to improve both the energy resolution and the neutron efficiency. In addition, this part is used as an active veto for external background. Indeed, the experimental site suffers a high background level of neutrons and $\gamma$, due to the surrounding neutron beam lines. That is why a heavy shielding made of B$_4$C, lead and borated polyethylene (HDPE) will surround the whole detector.

Cosmic muons are expected to constitute the main source of correlated background for the experiment. Indeed, high energy muons can produce fast neutrons by spallation in high-Z materials surrounding the detector. Therefore, a water-Cherenkov veto is placed above the detector to tag muon events. Reactor-off measurements will be used to determine background from undetected muons.

\section{Achievements}
\label{sec:Achievements}

The light and energy response of the detector has been validated by means of simulations reproducing the radioactive source measurements performed with a cell prototype. A homogeneous detector response with a 12\,\% energy resolution for 2\,MeV positrons is expected, as well as a  neutron detection efficiency of about 60\,\% assuming a 5\,MeV $\gamma$ threshold.

The detector has been built and is currently being tested using the light injection system which will be used to monitor the individual PMT properties as well as the cells optical properties. The detector stability between full calibrations is expected to be monitored at the percent level. Energy scale calibrations will be carried out by an automated radioactive source circulation around the detector vessel, ensuring a 2\,\% associated uncertainty.

Along 2014 and 2015, several campaigns of on-site measurements have allowed to characterize the STEREO site. Among them, the influence of a 15\,T superconducting magnet located in the neighbourhood of the experiment on the PMTs collection efficiency has been studied. A soft-iron -- mu-metal shielding was designed to reduce collection efficiency variations below the 0.3\,\% level for target PMTs. Concerning the background control, the main neutron and $\gamma$ sources have been identified and shielded by additional lead and HDPE walls combined with B$_4$C to mitigate the thermal neutron background. To further reject the neutron background, pulse shape discrimination is allowed by the scintillator properties and dedicated electronics \cite{Bourion2016} hosted in a $\mu$TCA crate. These electronics have been used to characterize the muon veto which exhibits a stable 99.5$\,$\% detection efficiency over time. On-site measurements using a cosmic wheel also permitted to validate the expected muon rate in the veto, computed using simulations. Combined together, the active and the passive shieldings should lower the background level down to the required signal to background ratio of 1.5.

With such a ratio and 400 detected $\bar{\nu_e}$ per day during 300 days, the STEREO experiment will be able to probe at 95\,\% C.L.  the reactor anti-neutrino anomaly within 2 years. The transportation of the detector inside the reactor building is imminent and will be followed by commissioning. Last shielding components will be installed this Summer and the first data taking is planned in early fall 2016.

\end{document}

%% file: econfmacros.tex
\def\beq{\begin{equation}}
\def\eeq#1{\label{#1}\end{equation}}
\def\eeqn{\end{equation}}

\def\beqa{\begin{eqnarray}}
\def\eeqa#1{\label{#1}\end{eqnarray}}
\def\eeqan{\end{eqnarray}}

\let\bar=\overbar

\def\etal{{\it et al.}}

\def\Dslash{\not{\hbox{\kern-4pt $D$}}}
\def\dslash{\not{\hbox{\kern-2pt $\del$}}}

\def\msb{{\bar{\ssstyle M \kern -1pt S}}}

